\documentstyle[12pt]{article}
\title{
{\normalsize
\begin{flushleft}
DESY 91--111
\\
PHE 91-10
\\
August 1991
\end{flushleft}
}
\vspace{1.5cm} \LARGE \bf
S-Matrix Approach to the Z Line Shape
\vspace{1.5cm}
}
\author
{
{\bf A. Leike$^{\ast}$, T. Riemann$^{\ast}$}\vspace{.3cm}\\
{
\it Institut f\"ur Hochenergiephysik, Platanenallee 6} \\
{\it O-1615 Zeuthen, Germany \vspace{.8cm}
}
\\and
\vspace{.8cm}
\\{\bf J. Rose$^{\ast}$} \vspace{.3cm}\\
{
\it III. Physikalisches Institut A der Rheinisch-Westf\"alischen Hochschule,
}
\\ {\it Sommerfeldstr. 26, W-5100 Aachen, Germany}
}
\date{August 8, 1991}

\newcommand{\z}{$Z\:$}

\begin{document}
\maketitle
\vspace{2cm}
\begin{abstract}
We analyze the \z line shape
assuming
the existence of an analytic, unitary S-matrix.
As an example, from hadron production at LEP we determine
$M_{Z}=91.134 \pm 0.020 \pm 0.020 \: ({\rm LEP})$ GeV,
$\Gamma_{Z}=2.506 \pm 0.018$ GeV.
This is in accordance with earlier results after performing
a shift of the $Z$ mass value of about
$\frac{1}{2}\Gamma_Z^2/M_Z = 34$ MeV.
The cross section and related observables may be described by a small
number of additional degrees of freedom
without relying on a specific field-theoretic model.
\end{abstract}
\vfill
\hrule width5cm
\vskip6pt
$^{\ast}$ {\small Partly supported by the German Bundesministerium f\"ur
Wissenschaft}
und
Technologie \hfill
\eject
\section*{}
In the early days of hadron physics, the contours of a satisfying dynamic
theory were far from obvious.
The analysis of resonance scattering had to be performed
with a minimum of theoretical assumptions.
Basics of the S-matrix theory which was developed in this context
and its application to the description
of resonances may be found e.g. in \cite{eden,sakurai,landau}.

The present understanding of the gauge theory of electroweak
interactions \cite{gws} allows for
detailed and precise theoretical predictions of electroweak scattering
including the precise measurements of \z boson interactions at LEP100.
Nevertheless, it is of some interest
to perform also model-independent fits to the \z line shape
\cite{usual,borrelli}.
In this respect, we consider the S-matrix theory to be the most consequent
approach. An introduction to the necessary formalism
and its application is the subject of the present article.
\vspace{.4cm}

The annihilation of electrons and positrons into
lepton pairs or hadrons at LEP100,
\begin{equation}
e^+e^-  \longrightarrow (\gamma, Z) \longrightarrow f^+f^-(\gamma),
\label{ee}
\end{equation}
is used to determine mass $M_Z$ and width $\Gamma_Z$ of the \z boson. These
observable
quantities correspond uniquely to the location of a pole of the S-matrix
describing
(\ref{ee}) in the complex energy plane:
\begin{equation}
{\bf \cal M}(s) = \frac {R_{\gamma}}{s} + \frac {R_{Z}}{s-s_{Z}} + F(s).
\label{smatrix}
\end{equation}
The poles of ${\bf \cal M}$ have complex residua
$R_{Z}$ and $R_{\gamma}$, the latter corresponding to the photon,
and $F(s)$ is an analytic function without poles. Further,
\begin{equation}
s_Z = M_Z^2 - i M_Z \Gamma_Z.
\label{s0}
\end{equation}
The analysis of the \z line shape
will be based here on the cross section
\begin{equation}
\sigma(s) = \sum_{i=1}^4 \sigma^i(s) = \frac{1}{4} \sum_{i=1}^4
            s |{\bf \cal M}^i (s)|^2,
\label{sigma}
\end{equation}
where the sum must be performed over four helicity amplitudes with different
residua
$R_Z^i$ and functions $F^i(s)$
\footnote{An application of the
S-matrix formalism to $e^+e^-$-annihilation has been proposed also in
\cite{stuart}. It is not pointed out there that one has to rely on helicity
amplitudes and a simple-minded application of the formulae
discussed there would fail.}.

Although we will not perform a field theoretic interpretation here,
for the reader's convenience the Born predictions of $R_{\gamma}$ and
$R_Z$ in terms of
vector and axial vector couplings are shown:
\begin{equation}
R_{\gamma}^B = \sqrt{\frac{4\pi}{3} c_f (1+\frac{\alpha_s}{\pi})  }
               Q_e Q_f \alpha(s),
\label{rgamma}
\end{equation}
\begin{eqnarray}
R_Z^{0,B}=R_Z(e_L^{-}e_R^{+}\rightarrow f_L^{-}f_R^{+})=c(v_e+a_e)(v_f+a_f),
    \nonumber \\
R_Z^{1,B}=R_Z(e_L^{-}e_R^{+}\rightarrow f_R^{-}f_L^{+})=c(v_e+a_e)(v_f-a_f),
    \nonumber \\
R_Z^{2,B}=R_Z(e_R^{-}e_L^{+}\rightarrow f_R^{-}f_L^{+})=c(v_e-a_e)(v_f-a_f),
    \nonumber \\
R_Z^{3,B}=R_Z(e_R^{-}e_L^{+}\rightarrow f_L^{-}f_R^{+})=c(v_e-a_e)(v_f+a_f).
\label{helirz}
\end{eqnarray}
In the standard theory,
\begin{equation}
c = \sqrt{\frac{4\pi}{3}c_f(1+\frac{\alpha_s}{\pi})}
    \frac{G_{\mu}}{\sqrt2} \frac{M_Z^2}{2\pi},
\hspace{1cm} a_f=\pm\frac{1}{2}, \hspace{1cm}
v_f=a_f(1-4|Q_f|\sin^2\vartheta_W),
\label{const}
\end{equation}
where $c_f$ is a possible color factor in case of hadron production.
The corresponding functions $F^{i}(s)$ vanish in Born approx\-imation,
$F^{i,B}(s)=0$.
In general, the $F^{i}(s)$ contain non-resonating radiative corrections.
More details on the realization of ansatz (\ref{smatrix}) in the standard
theory may be found in \cite{stuart}.

Instead referring to field theory, we parametrize the
cross section (\ref{sigma}) as follows:
\begin{equation}
\sigma(s) =  \sum_{A} \sigma_A(s), \hspace{1cm} A =
                 Z, \gamma, F,  \gamma Z,ZF, F\gamma,
\label{sign}
\end{equation}
with the contributions:
\begin{equation}
\begin{array}{rlrl}
\sigma_Z(s) =        &
\displaystyle{\frac{s r_Z}{|s-s_Z|^2} },
  &r_Z =                & \frac{1}{4} \sum |R_Z^i|^2,
\vspace{.4cm} \nonumber \\
\sigma_{\gamma}(s) = &
\displaystyle{\frac{r_\gamma}{s}} ,
&r_{\gamma} =         &  |R_{\gamma}|^2,
\vspace{.4cm} \nonumber \\
\sigma_F(s) =        &   s r_F(s),                 &
r_F(s) =             &  \frac{1}{4} \sum |F^i(s)|^2,
\vspace{.4cm} \nonumber \\
\sigma_{\gamma Z}(s)=& 2 {\rm Re}
\displaystyle{\frac {C_{\gamma}^{\ast} C_Z} {s-s_Z}}, &
C_{\gamma} =         & R_{\gamma}, \hspace{.5cm} C_Z = \frac{1}{4} \sum R_Z^i,
\vspace{.4cm} \nonumber \\
\sigma_{ZF}(s) =     &  2  {\rm Re}
\displaystyle{\frac {s C_{ZF}(s)} {s-s_Z}},         &
C_{ZF}(s) =          &\frac{1}{4} \sum R_Z^i F^{i\ast}(s),
\vspace{.4cm} \nonumber \\
\sigma_{F\gamma}(s) =& 2 {\rm Re} \left[ C_{\gamma}^{\ast} C_F(s)\right],&
C_F(s) =             &\frac{1}{4} \sum F^i(s).
\end{array}
\label{sigA}
\end{equation}

After making denominators real one remains with the following formula
for the line shape:
\begin{equation}
\displaystyle{\sigma (s) = \frac{R + (s - M_Z^2) I }{|s-s_Z|^2}
+ \frac{r_{\gamma}}{s} + r_0 + (s-M_Z^2) r_1 + \ldots}
\label{sigfin}
\end{equation}
Besides $M_Z, \Gamma_Z$, the real constants $R, I, r_0$ and $r_1$ are
introduced:
\begin{eqnarray}
R =&  M_Z^2 \left[  r_Z
     + 2 (\Gamma_Z / M_Z ) \left( \Im m C_R + M_Z \Gamma_Z \Re e (C'_R) \right)
     \right],
\nonumber\\
I =& r_Z + 2 \Re e C_R,
\nonumber\\
C_R(s) =& C_\gamma^{\ast} C_Z + s_Z C_{ZF}(s), \nonumber\\
r_0 =& M_Z^2 \left[ r_F -M_Z \Gamma_Z \Im m (r'_F) \right] +
      \Re e C_r - M_Z \Gamma_Z \Im m C'_r, \nonumber \\
r_1 =& r_F + M_Z^2 \left[ \Re e (r'_F) - (\Gamma_Z / M_Z ) \Im m (r'_F) \right]
      + \Re e C'_r, \nonumber \\
C_r(s) =& C_{\gamma}^{\ast} C_F(s) + C_{ZF}(s).
\label{fitconst}
\end{eqnarray}
The energy-dependent functions $C_{ZF}, C_F, r_F$,
and their (primed) derivatives with
respect to s have to be taken at $s=s_Z$. As may be seen, the cross section
may be described by only six real parameters as long as one takes into account
only the first two terms in the expansion of the functions $F^i(s)$ around
$s=s_Z$ and at most terms of the order $(s-M_Z^2)^n, n=0,1$ in the cross
section parametrization.

Next we have to discuss a conceptual problem due to QED bremsstrahlung.
Initial state radiation of photons leads to a deviation of the
{\it effective}
energy variable $s'$ in (\ref{sigma})
from $s$ with $s'<s$. At least events with soft photon
emission  unavoidably become part of the measured cross section. Taking them
into account means summing up an infinitely dense chain of single poles
leading to a new singularity structure (including a cut in the
complex plane) compared to the original
ansatz (\ref{sigma}). Indeed, the well-known formulae  for initial state
radiation in reaction (\ref{ee}) (see e.g. \cite{usual,h0t} and refs. therein)
contain in one way or the other the complex logarithm
\begin{equation}
L(s_Z) = \ln \frac{1-\Delta-s_Z/s}{1-s_Z/s},
\label{log}
\end{equation}
where $0 < \Delta < 1$ stands for some cut on the allowed photon energies.
A function like (\ref{log}) with its highly singular behaviour at $s=s_Z$
cannot be absorbed into the function $F(s)$ as introduced in (\ref{smatrix}).

The QED bremsstrahlung must be treated as follows. Initial state radiation
has
to be taken into account as exactly as possible,
e.g. using a convolution formula \cite{usual},
\begin{equation}
\sigma_T(s) = \int ds' \sigma (s') \rho_{\rm ini}(1-s'/s).
\label{convolution}
\end{equation}
Final state radiation can be
either calculated similarly or
formally simply
neglected. It doesn't influence the singularity structure of the
cross section and leads to some modifications of parameters other
than
$s_Z$ in (\ref{smatrix})
\footnote{If one wants to interpret the residua $R_{\gamma}$ and $R_Z$,
and $F(s)$ in terms of a field
theory, one should of course make an explicit calculation of final state
radiation.}.
The radiation connected with initial-final state
interferences can be taken into account by an analogue formula to
(\ref{convolution}) with
a slightly more complicated structure \cite{h0t,intconv}:
\begin{equation}
\sigma_{\rm int}(s) = \int ds' \sigma(s,s') \rho_{\rm int}(1-s'/s).
\label{convigif}
\end{equation}
The correct ansatz for the S-matrix based cross section is:
\begin{equation}
\sigma(s,s') = \frac{1}{8} s' \sum_i
               \left[ {\bf \cal M}_i(s) {\bf \cal M}_i^{\ast}(s')
                   +  {\bf \cal M}_i^{\ast}(s){\bf \cal M}_i(s') \right].
\label{sigif}
\end{equation}
We only mention that a representation like (\ref{sigA})
may be obtained easily also for $\sigma(s,s^{\prime})$.
If necessary, cross section (\ref{convigif}) may be added to
(\ref{convolution}).
Its numerical contribution is very small at LEP100 energies under usually
applied cut conditions.
\\
\vspace{.5cm}

Using (\ref{smatrix}) - (\ref{sigma}) for a fit to data, one is free of any
model-dependent
assumption, or
some choice of gauge, or a truncation of perturbation theory as must be usually
taken into account (see e.g. \cite{usual} and the recent discussion in
\cite{stuart,sirlin,spanier}).
The actual
configuration of cuts applied to the data is as unimportant as are details
of the final state. In case of a differential cross section, the S-matrix
would depend on additional variables.
\\
\vspace{.5cm}

In order to demonstrate that the S-matrix approach may have some practical
relevance, we use a simple code \cite{zpole} for the calculation of the QED
corrections
(\ref{convolution})
including soft photon exponentiation.
We perform four fits with a rising number of degrees of freedom to published
data at seven different beam energies $E$, $s=4E^2$, taken from
an analysis of the hadronic line
shape (Table 2 of \cite{l3}).
Without loss of generality, one can assume that the behaviour of the
running coupling constant of the photon $\alpha(s)$ is known
at LEP100 energies \cite{alfrun}:
\begin{equation}
\alpha \left( s=(91.2 {\rm GeV})^2 \right) = \alpha_0(1.0660-i0.0189).
\label{alfalep}
\end{equation}

For comparison, we give also the field theoretic Born estimates
for hadron production:
\begin{eqnarray}
R_{\gamma}^{B,h} &=& r^{1/2}
                     \sqrt{ \frac{4\pi}{3} c_f (1+\frac{\alpha_s}{\pi})}
                     \alpha(s),
\nonumber \\
r_Z^{B,h} &=& c^2 (v_e^2+a_e^2)
            [3(v_d^2+a_d^2)+2(v_u^2+a_u^2)]  \sim  6.32\ 10^{-4},
\nonumber \\
C_Z^{B,h} &=& \frac{c}{r^{1/2}} Q_e v_e [3 Q_d v_d + 2 Q_u v_u ]
          \sim 7.77 \ 10^{-4},
\nonumber \\
        r &=& 3 Q_d^2 + 2 Q_u^2,
\nonumber \\
C_{ZF}^{B,h} &=& C_F^{B,h} = r_F^{B,h} = 0.
\label{prediction}
\end{eqnarray}
Further,
\begin{eqnarray}
R^{B,h} =& M_Z^2 r_Z^{B,h} &= 5.45 \: {\rm GeV^2},
\nonumber \\
I^{B,h} =& r_Z^{B,h} + 2 (\Re e C_{\gamma}^{\ast}) C_Z &= 7.05 \times 10^{-3},
\nonumber \\
r_0^{B,h} =& 0.0 \: {\rm GeV^{-2}},
\nonumber \\
r_1^{B,h} =& 0.0 \: {\rm GeV^{-4}}.
\label{predic}
\end{eqnarray}

These numerical estimates are obtained with the weak parameters quoted in
\cite{l3}.

In our first fit with four free parameters we fix all quantities which are zero
in the Born approximation.
Then we allow for additional parameters ($r_0, r_1$) to be fitted.
The numerical results are shown in Table \ref{table}.
The small number of available data points is
certainly disadvantageous for the fit results.
Nevertheless, the table gives some impression on the potential
value of the approach. The gain of accuracy for a smaller number of
floating parameters
is a measure of the degree of biasing the fits with certain assumptions usually
done in a specific ansatz. From our starting point it is evident what would
be a completely unbiased fit - taking into account all higher powers of
$(s-M_Z^2)$ in the cross section ansatz and of $(s-s_Z)$ in the Taylor
expansions of the $F^i(s)$.
\begin{table}\centering
\begin{tabular}[h]{|c|*{4}{r@{$\pm$}l|}}
\hline
\multicolumn{1}{|c|}{ }   &\multicolumn{2}{c|}{ }  &\multicolumn{2}{c|}{ }
                          &\multicolumn{2}{c|}{ }  &\multicolumn{2}{c|}{ }  \\
nb. of
  &\multicolumn{2}{c|}{ }  &\multicolumn{2}{c|}{ }
  &\multicolumn{2}{c|}{ }  &\multicolumn{2}{c|}{ }  \\
parameters
  &\multicolumn{2}{c|}{4} &\multicolumn{2}{c|}{5}
  &\multicolumn{2}{c|}{5} &\multicolumn{2}{c|}{6}   \\
   \multicolumn{1}{|c|}{ }&\multicolumn{2}{c|}{ }  &\multicolumn{2}{c|}{ }
                          &\multicolumn{2}{c|}{ }  &\multicolumn{2}{c|}{ }  \\
\hline
\multicolumn{1}{|c|}{ }  &\multicolumn{2}{c|}{ }  &\multicolumn{2}{c|}{ }
                         &\multicolumn{2}{c|}{ }  &\multicolumn{2}{c|}{ }  \\
$M_{Z}$
  &91.134&.020   &91.130 & .020    &91.120& .032          &91.128 & .046
\\
$\Gamma_{Z}$
  & 2.506 & .018     & 2.484 & .040     & 2.490 & .034     & 2.484 & .041
\\
\multicolumn{1}{|c|}{ }  &\multicolumn{2}{c|}{ }  &\multicolumn{2}{c|}{ }
                         &\multicolumn{2}{c|}{ }  &\multicolumn{2}{c|}{ }  \\
\hline
\multicolumn{1}{|c|}{ }  &\multicolumn{2}{c|}{ }  &\multicolumn{2}{c|}{ }
                         &\multicolumn{2}{c|}{ }  &\multicolumn{2}{c|}{ }  \\
$R, {\rm GeV^{2}} $
  &5.49 & 0.08       & 5.38 & 0.21      & 5.41 & 0.17       & 5.38 & 0.21
\\
$I \times 10^{3}$
  &8.9 & 2.4         & 9.5 & 2.5        & 12.1 & 6.3        & 10.1 & 11.3
\\
\multicolumn{1}{|c|}{ }  &\multicolumn{2}{c|}{ }  &\multicolumn{2}{c|}{ }
                         &\multicolumn{2}{c|}{ }  &\multicolumn{2}{c|}{ }  \\
\hline
\multicolumn{1}{|c|}{ }  &\multicolumn{2}{c|}{ }  &\multicolumn{2}{c|}{ }
                         &\multicolumn{2}{c|}{ }  &\multicolumn{2}{c|}{ }  \\
$r_0 \times 10^7, {\rm GeV^{-2}}$
  &\multicolumn{2}{c|}{--}   & 3.5 & 5.9 &\multicolumn{2}{c|}{--}  & 3.0 & 13.1
\\
$r_1 \times 10^{10}, {\rm GeV^{-4}} $
  &\multicolumn{2}{c|}{--} &\multicolumn{2}{c|}{--}
  & --5.6 & 10.5        & --1.3 & 23.
\\
\multicolumn{1}{|c|}{ }  &\multicolumn{2}{c|}{ }  &\multicolumn{2}{c|}{ }
                         &\multicolumn{2}{c|}{ }  &\multicolumn{2}{c|}{ }  \\
\hline
\end{tabular}
\caption{Results of S-matrix based fits to the hadronic line shape
as measured at LEP100. An uncertainty of $20$ MeV in the energy scale of
LEP must yet be added to the error of $M_Z$.}
\label{table}
\end{table}

As is known from earlier fits, the determination of the residuum
and of the non-resonating terms is not too stringent if one analyzes only the
line shape. While our accuracy for the $Z$ width is comparable to other
determinations, we have a larger error for the mass. This is due to a strong
correlation between $M_Z$ and the parameters $I, r_1$ in (\ref{sigfin}),
which are not fixed here from the beginning. If one would assume them to be
known from other sources, the mass determination would be better. Similarly,
the $Z$ width is correlated with $R, r_0$. The smaller error of $R$
compared to that of $I$ leads to the relatively small error of $\Gamma_Z$
compared to that of $M_Z$.
\vspace{.4cm}

The measured \z mass value differs from earlier determinations
by a non-negligible shift. This is an immediate
consequence
of the S-matrix approach.
The parametrization of the Breit-Wigner resonance formula
for the \z peak as being inspired by perturbation theory
assumes usually (but {\it not necessarily} \cite{consoli,willenbrock})
an s-dependent width function $\bar{\Gamma}_Z(s)$
\cite{usual,l3,bbhvn,opal,aleph,delphi}.
In our notations, this would correspond to the following ansatz:
\begin{eqnarray}
{\bf \cal M}^i(s) = \frac {R_{\gamma}}{s} + \frac {\bar{R}_{Z}^i}
                       {s-\bar{s}_{Z}(s)} + \bar{F}^i(s),
\nonumber \\
\bar{s}_Z(s) = \bar{M}_Z^2 - i \bar{M}_Z \bar{\Gamma}_Z(s).
\label{sbarmatrix}
\end{eqnarray}
The difference between $M_Z, \Gamma_Z, R_{Z}^i$ and $\bar{M}_Z,
\bar{\Gamma}_Z, \bar{R}_{Z}^i$ is described by a transformation
proposed earlier in another context \cite{ztrafo}:
\begin{eqnarray}
\bar{M}_Z = M_Z \sqrt{1+\Gamma_Z^2/M_Z^2}
    \approx M_Z + \frac{1}{2} \Gamma_Z^2/M_Z
    = M_Z+34 {\rm MeV}, \nonumber
\label{mbar}
\end{eqnarray}
\begin{eqnarray}
\bar{\Gamma}_Z = \Gamma_Z \sqrt{1+\Gamma_Z^2/M_Z^2}
    \approx \Gamma_Z + \frac{1}{2} \Gamma_Z^3/M_Z^2
    = \Gamma_Z + 1 {\rm MeV}, \nonumber
\label{gbar}
\end{eqnarray}
\begin{eqnarray}
\bar{R}_Z = R_Z (1 + i \Gamma_Z/ M_Z).
\label{zz}
\end{eqnarray}
This transformation
is exact as long as there are no thresholds (opening new
decay channels) or rapidly changing radiative corrections in the vicinity of
the \z peak position. Then,
\begin{equation}
\bar{\Gamma}_Z(s) = \frac {s} {\bar{M}_Z^2} \bar{\Gamma}_Z.
\label{gammas}
\end{equation}
If (\ref{gammas}) would be exact, it would follow $\bar{F}(s)=F(s)$.
A dependence of mass and width determinations on
the
theoretical ansatz for a line shape description has been observed earlier, see
e.g. \cite{sakurai}. There, formulae (\ref{zz}) may also be applied in order
to
relate different approaches to the hadron resonances under discussion.
We further mention that the ratio of width and mass is invariant:
\begin{equation}
\frac{ \bar{\Gamma}_Z} {\bar{M}_Z} = \frac{ \Gamma_Z} {M_Z}.
\label{wm}
\end{equation}
\vspace{.5cm}
\\

Although we think that the natural application  of the S-matrix approach
to \z boson physics is the line shape analysis, we indicate also how
other observables than $\sigma_T(s)$ may be treated.
For a calculation of e.g. the initial state bremsstrahlung contribution to the
left-right asymmetry $A_{LR}$,
the convolution for the numerator $\sigma_{LR}(s)$ must be performed with
a modified ansatz:
\begin{equation}
A_{LR} = \frac{\sigma_{LR}(s)}{\sigma_T(s)}, \hspace{1cm}
\sigma_{LR}(s) =
\int ds' \left[ \sigma_0 + \sigma_1 - \sigma_2 - \sigma_3 \right] (s')
\rho_{\rm ini}(1-s'/s).
\label{alr}
\end{equation}
The helicity cross sections $\sigma_i$ are introduced in (\ref{sigma}).
Similarly, the final state polarisation
$A_{pol}$ may be obtained:
\begin{equation}
A_{pol} = \frac{\sigma_{pol}(s)}{\sigma_T(s)}, \hspace{1cm}
\sigma_{pol}(s) =
\int ds' \left[ \sigma_0 - \sigma_1 - \sigma_2 + \sigma_3 \right] (s')
\rho_{\rm ini}(1-s'/s).
\label{apol}
\end{equation}
The forward-backward asymmetry $A_{FB}$
deserves additional comments.
Due to the different angular
integrations, the weight functions (flux factors) $R_{\rm ini}(z)$ etc.
of the forward-backward difference cross section $\sigma_{FB}(s)$
differ from those for $\sigma_T(s)$. Nevertheless, a convolution
may be derived \cite{h0t,intconv,ringberg}:
\begin{equation}
A_{FB} = \frac{\sigma_{FB}(s)}{\sigma_T(s)}, \hspace{1cm}
\sigma_{FB}(s) =
\int ds' \left[ \sigma_0 - \sigma_1 + \sigma_2 - \sigma_3 \right] (s')
R_{\rm ini}(1-s'/s).
\label{afb}
\end{equation}
Strictly speaking, the $\sigma_i$ for the forward-backward asymmetry are
different from those for the total cross section.
They agree in Born approximation. Further, we know from one-loop calculations
in the standard theory
which changes are to be expected after the introduction of e.g. weak
form factors as proposed in \cite{realistic}.

A study of the usefulness of these formulae should be performed with
a more sophisticated code than ZPOLE in order to describe more realistic
cut situations. A modified version of ZFITTER \cite{alfrun} for this purpose
is in preparation. A larger number of experimental
data points would also be highly desirable
due to the large number of degrees of freedom of the line shape.
\vspace{.8cm}\\

To summarize, we formulate an alternative approach to the \z line shape
assuming the existence of an analytic, unitary S-matrix and the validity
of QED.

We demonstrate that
the approach allows reasonable fit results for mass and width of the
\z boson. The \z mass obtained this way
differs by a well-understood shift of -34 MeV from earlier measurements.
The other line shape parameters determined by the fit can also be
interpreted e.g. in the standard theory.

A dedicated application of the S-matrix approach
to polarized scattering and to $Z'$ physics would also be
interesting.
\vspace{1.cm}
\nopagebreak
\begin{center}
{\bf Acknowledgements}
\end{center}
\vspace{.3cm}
We would like to thank D. Bardin,
A. B\"ohm, W. Hollik, W. Lohmann, R. Stuart, B. Ward
for interesting and stimulating
discussions
and W. Lohmann additionally for support.

\end{document}